\documentclass[a4paper,twocolumn,showpacs,superscriptaddress,floatfix]{revtex4-1}
\usepackage{graphicx}
\usepackage{epsf}
\usepackage{epsfig}
\usepackage{amssymb,amsmath}
\usepackage[usenames]{color}
\usepackage{amssymb}
\usepackage{times}

\newcommand{\ie}{\textit{i.e.}~}
\newcommand{\eg}{\textit{e.g.}~}

\begin{document}

\title{Analytic modelling of tidal effects in the relativistic inspiral of
  binary neutron stars}

\author{Luca \surname{Baiotti}}
\affiliation{Institute of Laser Engineering, Osaka University, Suita, Japan}
\affiliation{Yukawa Institute for Theoretical Physics, Kyoto University, Kyoto, Japan}

\author{Thibault \surname{Damour}}
\affiliation{Institut des Hautes Etudes Scientifiques, Bures-sur-Yvette, France}
\affiliation{ICRANet, Pescara, Italy}

\author{Bruno \surname{Giacomazzo}}
\affiliation{Department of Astronomy, University of Maryland, College Park, Maryland USA}
\affiliation{Gravitational Astrophysics Laboratory, NASA Goddard Space Flight Center, Greenbelt, Maryland USA}
\affiliation{Max-Planck-Institut f\"ur Gravitationsphysik, Albert-Einstein-Institut, Potsdam, Germany}

\author{Alessandro \surname{Nagar}}
\affiliation{Institut des Hautes Etudes Scientifiques, Bures-sur-Yvette, France}

\author{Luciano \surname{Rezzolla}}
\affiliation{Max-Planck-Institut f\"ur Gravitationsphysik, Albert-Einstein-Institut, Potsdam, Germany}
\affiliation{Department of Physics and Astronomy, Louisiana State University, Baton Rouge, Louisiana, USA}

\begin{abstract}
To detect the gravitational-wave (GW) signal from binary neutron stars and
extract information about the equation of state of matter at nuclear
density, it is necessary to match the signal with a bank of accurate
templates. We present the two longest (to date)
general-relativistic simulations of equal-mass binary neutron stars
with different compactnesses, ${\cal C}=0.12$ and ${\cal C}=0.14$, and
compare them with a tidal extension of the effective-one-body (EOB)
model.  The typical numerical phasing errors over the $\simeq 22$ GW
cycles are $\Delta \phi\simeq \pm 0.24$ rad.  By calibrating only one parameter (representing a
higher-order amplification of tidal effects), the EOB model can 
reproduce, within the numerical error, the two numerical waveforms 
essentially up to the merger.  By contrast, the
third post-Newtonian Taylor-T4 approximant with leading-order
tidal corrections dephases with respect to the numerical waveforms by
several radians.
\end{abstract}


\maketitle

\emph{Introduction.~}Inspiralling binary neutron stars (BNSs)
are among the strongest sources of gravitational waves (GWs) and
certain targets for the advanced and new-generation ground-based GW
detectors LIGO/Virgo/GEO/ET~\cite{Andersson:2010}. These detectors
will be sensitive to the inspiral GW signal up to GW frequencies of about 
$1$~kHz, which are reached just before the merger. The late inspiral
signal will be influenced by tidal interaction between the two neutron
stars (NSs), which, in turn, encodes important information about the
equation of state (EOS) of matter at nuclear densities. However, to
reliably extract such information, both a large sample of numerical
simulations and an analytical model of inspiralling BNSs which is able
to reproduce them accurately, are needed. In this \textit{Letter} we
report on significant progress on this problem by presenting the
longest (to date) simulations of merging equal-mass BNSs and by
showing how to use them to calibrate an effective-one-body (EOB) model
of tidally interacting BNSs.

Numerical simulations of merging BNSs in full general relativity 
have reached a high-level of accuracy and have become more realistic
(\eg, with the inclusion of magnetic fields) only 
recently~\cite{Baiotti08,Yamamoto2008,Anderson2008,Read:2009b}.
The analytical description
of tidally-interacting binary systems has been initiated very
recently~\cite{Flanagan:2007ix,Hinderer08,Hinderer:2009ca,Damour:2009wj},
with the following two major results. First, the dimensionless quantity
$k_\ell$ (Love number) in the (gravito-electric) tidal polarizability
parameter $G\mu_\ell \equiv 2 k_\ell R^{2\ell +1}/(2\ell -1)!!$
measuring the relativistic coupling (of multipolar order $\ell$)
between a NS of radius $R$ and the external gravitational field in
which it is embedded has been found to be a {strongly decreasing}
function of the compactness parameter ${\cal C}\equiv GM/(c^2 R)$ of
the NS. Second, a comparison between the numerical computation of the
binding energy of quasi-equilibrium circular sequences of
BNSs~\cite{Uryu:2009ye} and the EOB description of tidal
effects~\cite{Damour:2009wj} has suggested that higher-order
post-Newtonian (PN) corrections to tidal effects effectively increase 
by a factor of order two the tidal polarizability of close NSs. 
The main aim of this paper is to extend the domain of applicability of the EOB
method~\cite{Buonanno:1998gg}, from the inspiralling binary black hole
(BBH) case (for which it recently provided a very accurate analytic
description~\cite{Damour:2009kr,Buonanno:2009qa}), to the yet
unexplored case of inspiralling BNSs. To this aim we have performed two
accurate and long-term BNS simulations covering approximately $20-22$ GW cycles
of late inspiral, and we will show that they can be reproduced
accurately almost up to the merger by a new tidal extension of 
the EOB model.

\emph{Tidal corrections in the EOB approach.~}We
recall that the EOB formalism~\cite{Buonanno:1998gg} replaces the
PN-expanded two-body dynamics by a \textit{resummed} description with,
in particular, a Hamiltonian of the form: $H_{\rm EOB} \equiv M
c^2\sqrt{1+2\nu (\hat{H}_{\rm eff}-1)}$, where $M\equiv M_A + M_B$ is
the total mass and where \hbox{$\nu \equiv M_A \, M_B / (M_A +
  M_B)^2$} is the symmetric mass ratio. Here the ``effective
Hamiltonian'' $\hat{H}_{\rm eff}$ is a simple function of the momenta
and it incorporates the relativistic gravitational attraction mainly
through the so-called ``EOB radial potential'' $A(r)$. The structure
of $A(r)$ is remarkably simple at 3PN: $A^{\rm 3PN} (r) = 1-2u+2 \,
\nu \, u^3 + a_4 \, \nu \, u^4$, where $a_4 = 94/3 - (41/32)\pi^2$,
and $u \equiv GM/(c^2r_{AB})$. An excellent description of BBHs
has been found to be given by~\cite{Damour:2009kr}
%
$
A^0(r) = P^1_5\left[1-2u+2\nu u^3 + a_4 \nu u^4 + a_5\nu u^5 + 
a_6\nu u^6\right]  ,
$
%
where $P^n_m$ denotes an $(n,m)$ Pad\'e approximant and where values
of the coefficient $a_5=-6.37$, $a_6=+50$ provide a very good
agreement between EOB and numerical-relativity (NR) waveforms for
BBHs~\cite{Damour:2009kr} (the results presented here are 
insensitive to this choice as long as $a_5$ and $a_6$ are chosen 
in a well defined range). 
Ref.~\cite{Damour:2009wj} suggested to include tidal effects as 
corrections both to the radial potential and to the waveform (and
radiation reaction). The tidally corrected radial potential reads
$A(u) = A^0(u) + A^{\rm tidal} (u)$, where
\begin{align}
\label{eq:T9}
A^{\rm tidal}=\sum_{\ell\geq 2} - \kappa_\ell^T u^{2\ell+2}\hat{A}^{\rm tidal}_\ell(u)\,.
\end{align}
Here $\kappa^T_\ell u^{2\ell +2}$ describes the leading-order (LO) tidal
interactions. It is a function of the two masses, of the two
compactnesses ${\cal C}_{A,B}$, and of the two (relativistic) Love
numbers $k_\ell^{A,B}$
\begin{eqnarray}
\label{eq:def_kT}
\kappa_{\ell}^{ T} &=& 2 \dfrac{M_B \, M_A^{2\ell}}{(M_A + M_B)^{2\ell + 1}}
\dfrac{k_\ell^A}{{\cal C}_A^{2\ell + 1}} +  
\left\{ ~_A~\leftrightarrow~_B\right\}\,.
\end{eqnarray}
The additional factor $\hat{A}^{\rm tidal}_\ell(u)$ in
Eq.~\eqref{eq:T9} represents the effect of higher-order relativistic
contributions to the tidal interactions: next-to-leading order (NLO),
next-to-next-to-leading order (NNLO), etc.  A number of different
prescriptions can be considered for the correcting tidal factor 
$\hat{A}^{\rm  tidal}_\ell$ and these will be presented in a longer 
companion work~\cite{Baiotti:2010}. Here, we will limit ourselves 
to an $\ell$-independent, ``Taylor-expanded'' expression 
$\hat{A}^{\rm tidal}_\ell(u)\equiv 1
+\bar{\alpha}_1u + \bar{\alpha}_2u^2$~\cite{Damour:2009wj}, 
where $\bar{\alpha}_n$ are pure numbers in the equal-mass case, 
but functions of $M_A$, ${\cal C}_A$, and $k_\ell^A$ in the general
case. The analytical value of the ($\ell=2$) 1PN coefficient 
$\bar{\alpha}_1$ has been reported in~\cite{Damour:2009wj} 
(and recently confirmed in~\cite{Vines:2010ca}).
In the equal-mass case, it yields $\bar{\alpha}_1=1.25$.
We will use this analytical value in the following and 
use our simulations to constrain the value of the 
2PN coefficient $\bar{\alpha}_{2}$.
Similarly, one takes into account an $\ell=2$ tidal correction
to the waveform and radiation reaction, as given at LO in Sec.~V 
of~\cite{Damour:2009wj}. Additional coefficients parametrizing 
higher-order tidal relativistic contributions in the waveform 
and radiation reaction (such as $\beta_1$ in Eq.~(71) 
of~\cite{Damour:2009wj}), were found to have a small 
effect~\cite{Baiotti:2010} and will be neglected here.
In principle, tidal effects can also be accounted for via 
modifications of one of the \textit{non-resummed} PN models, 
such as the Taylor-T4 one; see below for its comparison with the 
NR results.

In order to measure the influence of tidal effects, it is useful to
consider the ``phase acceleration'' $\dot\omega \equiv d \, \omega /
dt \equiv d^2 \phi / dt^2$, where $\phi\equiv \phi_{22}$ is the
phase of  either the curvature or of the metric $\ell=m=2$ GWs. 
The function $\dot\omega(\omega)$ is independent of the two ``shift ambiguities''
that affect the GW phase $\phi (t)$, namely the shifts in time and
phase, and thus a useful intrinsic measure of the
quality of the waveform~\cite{DamourNagar:07b}. 
However, here we use another diagnostic to measure the phase acceleration, 
namely the dimensionless function
\begin{equation}
\label{eq:5.15}
Q_{\omega} (\omega) = \frac{d\phi}{d \, \ln \, \omega} =
\frac{\omega \, d \phi / dt}{d\omega / dt} =
\frac{\omega^2}{\dot\omega} 
 \, .
\end{equation}
%

\begin{figure}[t]
\begin{center}
\includegraphics[width=8.0cm,angle=0]{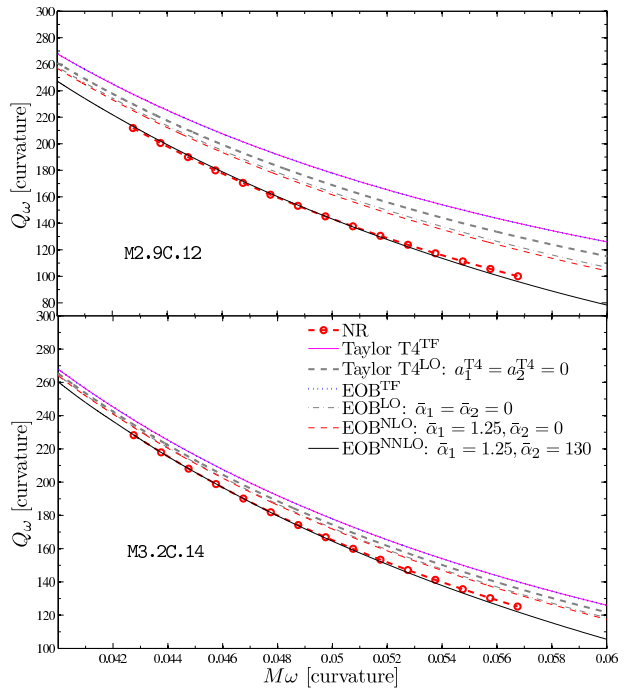}
\caption{\label{fig:fig1} Comparison of the EOB $Q_\omega$  
curves for different choices of the effective tidal amplification 
factor $\hat{A}_\ell^{\rm tidal}(u) = 1+\bar{\alpha}_1 u +\bar{\alpha}_2 u^2$,
with the corresponding NR ones (dashed lines with open circles) 
for the two binaries considered. The dotted line corresponds to
the ``tidal-free'' ( or ``point-mass'') EOB, 
namely when ignoring tidal effects. 
The figure also includes two Taylor-T4 models: tidal-free, 
and augmented by LO tidal effects.}
  \end{center}
\end{figure}

\emph{Numerical Simulations.~}They were performed
with the {\tt Cactus}-{\tt Carpet}-{\tt
  Whisky}~\cite{Schnetter-etal-03b} codes and, in essence, 
we use the same gauges and numerical methods already discussed
in~\cite{Baiotti08}. 
As initial data we use quasi-equilibrium irrotational binaries generated
with the multi-domain spectral-method code \texttt{LORENE}, within a
conformally-flat spacetime metric~\cite{Gourgoulhon01}. The EOS of the
initial data is the polytropic one $p = K\,\rho^\Gamma$, where $p$,
$\rho$, $K = 123.6$, and ${\Gamma = 2}$ are the pressure, rest-mass
density, polytropic constant, and adiabatic index, respectively (in
units where $c=G=M_\odot=1$). 
The evolutions are instead performed
with either a polytropic EOS or an ``ideal-fluid'' one, $p=\rho
\epsilon (\Gamma-1)$, where $\epsilon$ is the specific internal
energy; the differences in phasing introduced by the different 
EOSs are of order $\pm 0.13$ rad~\cite{Baiotti:2010}. 
Because the stellar compactness represents the
most important parameter determining the size of tidal effects, we
have considered two different (equal-mass) binaries having total 
Arnowitt-Deser-Misner~(baryonic) mass of either 
$2.69$~($ 2.89)\,M_\odot$ or $3.00$~($3.25$)$\,M_\odot$, 
thus with compactnesses ${\cal C}\equiv{\cal C}_A= {\cal C}_B=0.12$ 
or ${\cal C}=0.14$. 
The dominant ($\ell=2$) tidal parameters for the 
two compactnesses ${\cal C}=0.12$ $(0.14)$ are found 
to be~\cite{Hinderer08}, respectively,
$k_2\equiv k_2^A= k_2^B=0.00969~(0.07894)$, 
and therefore $\kappa^T_2=496.01~(183.81)$.
Hereafter the two binaries will be referred to as \texttt{M2.9C.12} and
\texttt{M3.2C.14}, respectively. The number of refinement levels and
their resolutions are the same as those in~\cite{Baiotti08}, but the
initial coordinate separation between the stellar centers is 
$60$~km, \ie considerably larger than the one considered in~\cite{Baiotti08}. 
This yields about $10$ orbits before merger, thus
the two longest BNS waveforms produced to date.

\begin{figure*}[t]
\begin{center}
\includegraphics[width=8.0cm,clip=true]{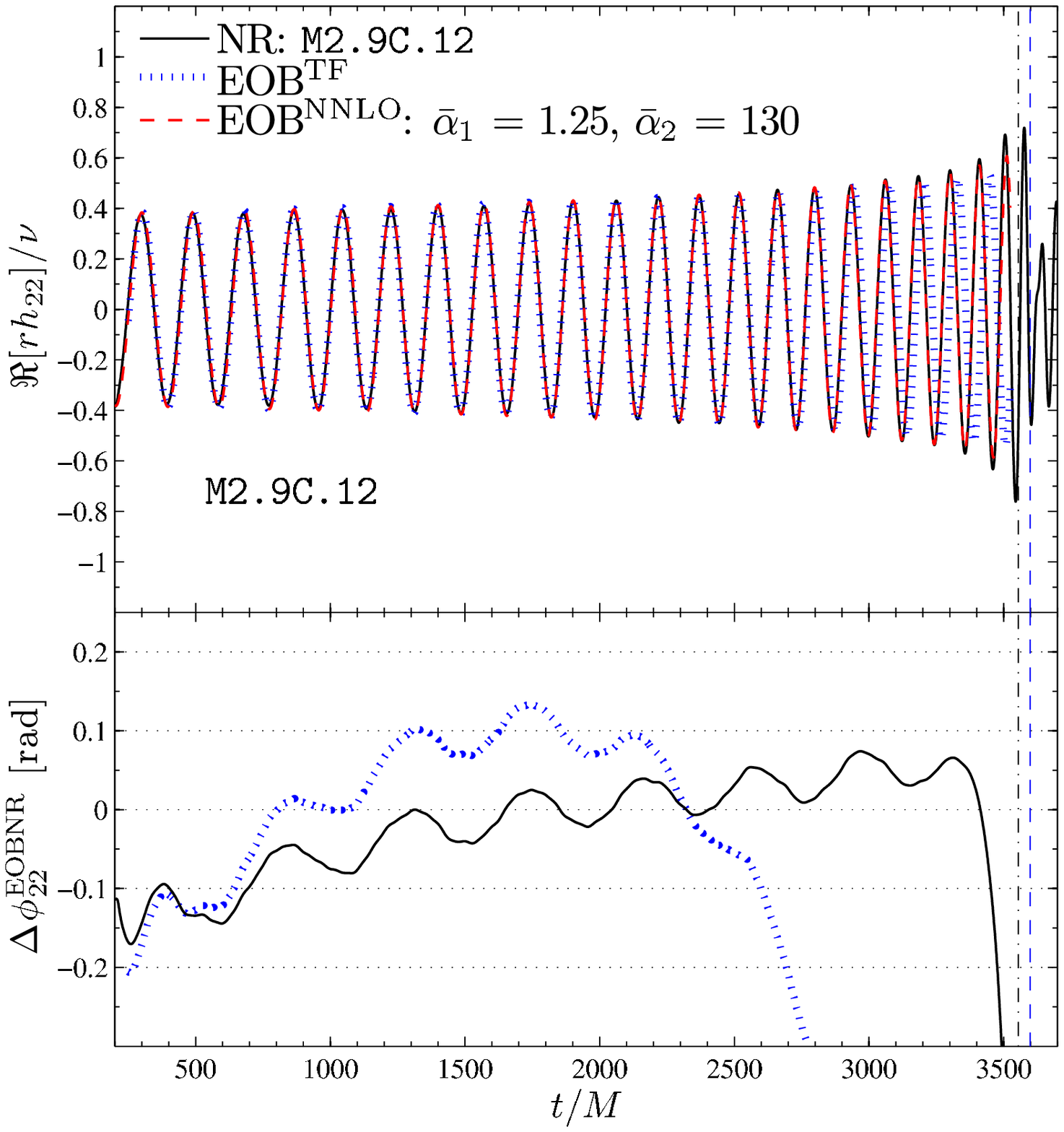}
\hskip 1.0cm
\includegraphics[width=8.0cm,clip=true]{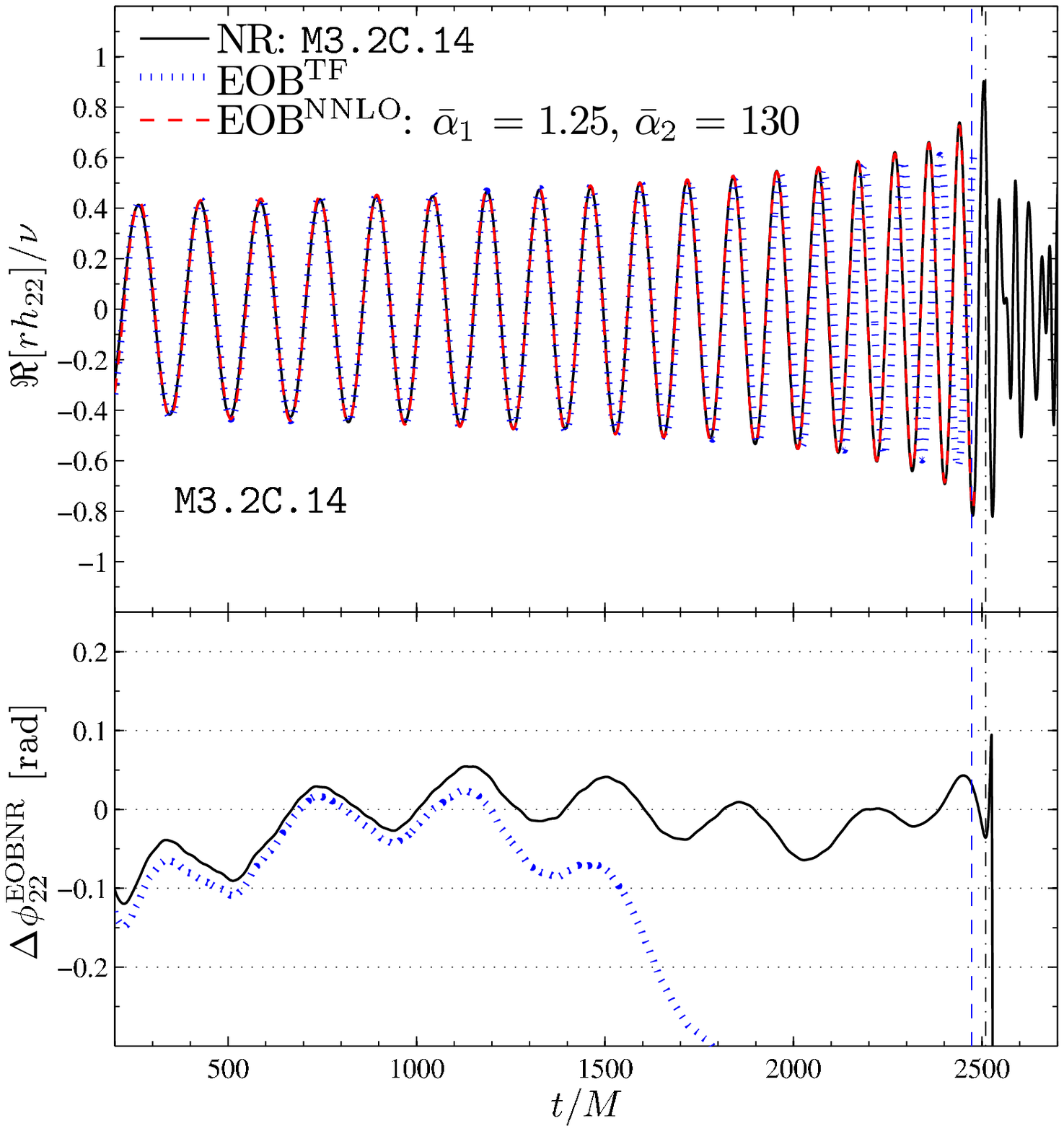}
\caption{\label{fig:fig3}Comparison between NR and EOB phasing for the
  {\tt M2.9C.12} (left panels) and {\tt M3.2C.14} (right panels)
  binaries. The top panels show the real parts of the $h_{22}$
  waveforms, while the bottom panels show the corresponding phase
  differences. Note the excellent agreement almost up to the time of
  merger (vertical dashed and dot-dashed lines) and the very large
  errors when tidal effects are neglected (dotted line).}
\end{center}
\end{figure*}

\emph{Discussion.~}We start our comparison between
the NR results and the EOB ones by showing in Fig.~\ref{fig:fig1} 
the $Q_\omega$ diagnostics for various possible
LO/NLO tidal models and for scaled GW frequencies $M \omega \lesssim 0.06$
[\ie up to $3~(5)$ GW cycles before merger for the 
{\tt  M2.9C.12} ({\tt M3.2C.14}) binary]. 
[We have estimated the NR $Q_\omega$ by fitting each 
NR phase evolution on a (scaled) frequency interval $I=[0.043, 0.057]$ 
with an analytical expression that reproduces at lower order the behavior
expected from the PN approximation, thereby filtering out the amplified 
numerical noise coming from the two time derivatives in $\dot{\omega} \equiv d^2 \phi / dt^2$
(more details will be presented in~\cite{Baiotti:2010})].
The first thing to note is that both the {\it tidal-free} 
EOB model  (EOB$^{\rm TF}$, dotted line) and the EOB model
including only LO tidal corrections (EOB$^{\rm LO}$, dot-dashed line) 
are clearly unable, both for the {\tt M2.9C.12} (upper panel) and the 
{\tt  M3.2C.14} binaries (lower panel), to match the corresponding 
NR curves (dashed line with open circles). 
The dephasing accumulated over the frequency interval $I$ 
%
$
\Delta_I\phi^{\rm EOBNR}\equiv\int_I(Q_\omega^{\rm EOB}-Q_\omega^{\rm NR}) d\ln \omega \,,
$
%
by the EOB$^{\rm LO}$ model relative to the 
${\cal C}=0.12~(0.14)$ NR data is about $5.5~(2.0)$ rad.  
This is much larger than the NR phasing error, related to a 
finite-radius extraction and EOS dependence, 
estimated to be $\Delta\phi=\pm 0.24$~\cite{Baiotti:2010}.

The inclusion of the NLO 1PN tidal effect 
($\bar{\alpha}_1=1.25$~\cite{Damour:2009wj}) only slightly 
reduces these dephasings to about $4.9~(1.8)$ rad (EOB$^{\rm NLO}$ curves 
in Fig.~\ref{fig:fig1}).
This clearly indicates the need for NNLO (2PN and higher) tidal effects.
We then found that choosing  $\bar{\alpha}_2\approx 130$, yields
a good match between the $Q_\omega$ curves (solid line, EOB$^{\rm NNLO}$) 
and the NR data (dashed line with open circles) for
\textit{both} binaries, with a corresponding 
phase difference $\Delta_I\phi^{\rm EOBNR}$~$\approx 0.1$ rad.
The value $\bar{\alpha}_2 \approx 130$ is probably only an 
effective description of higher-order relativistic tidal effects. 
Moreover, the precise value of $\bar{\alpha}_2$, or more generally of 
the amplification factor $\hat{A}^{\rm tidal}_\ell (u)$, is sensitive 
to the numerical truncation error. When considering 
resolution-extrapolated GWs~\cite{Baiotti:2010},  we found 
a smaller value of $\bar{\alpha}_2$, which is compatible 
with the estimate obtained using the binding energy of 
circular BNSs~\cite{Uryu:2009ye}.

Figure~\ref{fig:fig1} also reports the $Q_\omega$ diagnostics 
obtained when using two versions of the  Taylor-T4 approximant:
the tidal-free model (T4$^{\rm TF}$, magenta, upper solid line), 
and the Taylor-T4$^{\rm LO}$ one (thick-dashed line).
The latter model includes only the LO tidal effects~\cite{Flanagan:2007ix}, 
\ie the LO tidal contribution $a^{\rm tidal}(x)\propto \kappa_2^T x^5$ to $dx/dt$
[where $x\equiv(M\omega/2)^{2/3}$; see~\cite{Hinderer:2009ca} and Eqs.~(86)--(88) of~\cite{Damour:2009wj}].
Note that the tidal-free Taylor-T4 $Q_\omega$ curve nearly coincides
with the tidal-free EOB one, with a dephasing $\Delta_I\phi^{\rm T4EOB}=0.013$ rad. 
On the other hand, the $I$-integrated dephasings between the T4$^{\rm LO}$ description 
and the NR results are rather large, namely $\Delta_I\phi^{\rm T4NR}=6.96~(2.53)$
rad  
for ${\cal C}=0.12~(0.14)$. We have investigated whether a suitable 
PN-amplification factor  $\hat{a}^{\rm tidal}(x)=1+a_1^{\rm T4}x+a_2^{\rm T4}x^2$
of $a^{\rm tidal}(x)$ might accurately reproduce the NR data, but we found 
that this was not possible with a \textit{single} choice of 
$\hat{a}^{\rm tidal}(x)$ for the two simulations~\cite{Baiotti:2010}.
The latter result suggests that the EOB modelling of tidal effects 
may be more robust than the corresponding Taylor-T4 one

We next consider the comparison of the waveforms in the time domain
and \textit{over the full inspiral up to the merger}. This is shown in
Fig.~\ref{fig:fig3}, whose left panels refer to the {\tt M2.9C.12}
binary and the right ones to {\tt M3.2C.14}. The top parts
compare the (real part) of the EOB$^{\rm NNLO}$ 
(with $\bar{\alpha}_1=1.25$, $\bar{\alpha}_2=130$) 
and NR metric $h_{22}$ waveforms, while the bottom panels
show the corresponding phase differences, $\Delta\phi^{\rm EOBNR}(t)
\equiv \phi^{\rm EOB}(t)-\phi^{\rm NR}(t)$ (suitably shifted in time
and phase \`a la~\cite{Damouretal:2007}). The two vertical lines
indicate two possible markers of the ``time of the merger''; more
specifically, the dashed lines refer to the NR merger, defined as
the time at which the modulus of the metric NR waveform reaches 
its first maximum, while the vertical dash-dotted line represents 
the EOB estimate of the ``formal'' contact~\cite{Damour:2009wj}.
Figure~\ref{fig:fig3} clearly shows that the agreement in the time
domain between the analytic EOB description and the numerical one is
extremely good essentially up to the merger, with a phase error 
which is well within the estimated error bar. 
More precisely: (i) in the  {\tt M3.2C.14} case, the phase difference 
$\Delta\phi^{\rm EOBNR}(t)$ varies between $-0.1$ rad in the early
inspiral and $+0.04$ rad at the NR merger; (ii) in the  
{\tt M3.2C.12} case, $\Delta\phi^{\rm EOBNR}(t)$ varies 
between $-0.15$ rad in the early inspiral and $-0.15$
rad as late as $130M$ (\ie approximately $1.5$ GW cycles) before
the NR merger. For the latter binary, it is only just before the 
NR merger that the dephasing becomes of order 1 rad.

\emph{Conclusions.~}We have presented the first
NR-EOB comparison of the GWs emitted during the inspiral of
relativistic BNSs. In particular, we have analyzed the longest to date
numerical simulations of equal-mass, irrotational BNSs with two
different compactnesses. We found that tidal effects are 
{\it significantly amplified by higher-order relativistic corrections} 
(2PN and higher) and that the inclusion of such corrections is necessary for an
accurate phasing of the GW signal. 
Such an amplification is equivalent to a (distance-dependent) 
{\it effective increase} of the Love numbers.  
When a \textit{single} choice for the {\it  unique} free parameter 
in the NNLO term is made, the EOB model is
able to reproduce the two NR phase evolutions well within the
estimated NR error and essentially up to merger (and in fact up to
merger in the ${\cal C}=0.14$ case).  By contrast, we have shown that
the use of the Taylor-T4 PN approximant considered
in~\cite{Hinderer:2009ca} leads to phase disagreements 
(over  the frequency interval $I=[0.043,0.057]$) 
$\Delta_I\phi^{\rm T4NR}= 6.96~(2.53)$ rad for ${\cal C}=0.12~(0.14)$.

The work reported here provides the first evidence that an accurate
analytic modelling of the late inspiral of tidally interacting BNSs is
possible, thereby opening the possibility to extract reliable
information on the EOS of matter at nuclear densities from the data of
the forthcoming advanced GW detectors. These encouraging results,
however, also call for a continued synergy between more accurate
numerical simulations (notably exploring different mass ratios) 
and higher-order analytic results.


\smallskip\noindent\emph{Acknowledgments.~} We thank the developers of
                        {\tt Cactus} and {\tt Carpet} for their
                        continuous improvements. The simulations were
                        performed on Ranger (TACC/TG-MCA02N014) and
                        Damiana (AEI). This work was supported by the
                        DFG grant SFB/Transregio~7, by ``CompStar'', a
                        Research Networking Programme of the ESF, by
                        the JSPS Grant-in-Aid for Scientific Research
                        (19-07803), by the MEXT Grant-in-Aid for Young
                        Scientists (22740163), and by NASA
                        (NNX09AI75G).


\end{document}